\documentclass[10pt, journal, twocolumn]{IEEEtran}

\usepackage[utf8]{inputenc}
\usepackage[scr=boondoxo,frak=euler,bb=ams]{mathalfa}
\usepackage{amsmath,amsthm,amsfonts,amssymb}
\usepackage{graphicx,graphics}
\usepackage{float}
\usepackage{tikz}
\usetikzlibrary{shapes, snakes, patterns}
\usepackage{xcolor}

\newcommand{\spheading}[2][5em]{% 
	\rotatebox{90}{\parbox{#1}{\raggedright #2}}}
\usepackage[export]{adjustbox}
\usepackage[justification=centering]{caption}
\usepackage{subcaption}
\DeclareCaptionLabelSeparator{periodspace}{.\quad}
\usepackage{epstopdf}
\usepackage{enumerate,enumitem} 
\usepackage{cite}
\usepackage{suffix}
\usepackage{mathtools}
\usepackage{tabularx}
\usepackage{booktabs}
\usepackage{boldline,multirow,diagbox,hhline} 
\usepackage{slashbox}
\usepackage{algorithm,algpseudocode}
\usepackage{relsize}
\usepackage{cuted}
\usepackage{titlesec}
\usepackage{chngcntr}
\usepackage{textcomp}
\usepackage{balance}
\def\BibTeX{{\rm B\kern-.05em{\sc i\kern-.025em b}\kern-.08em
		T\kern-.1667em\lower.7ex\hbox{E}\kern-.125emX}}
        
\usepackage{stfloats} % <---
\theoremstyle{definition}

\newtheorem{theorem}{Theorem}[]
\newtheorem{lemma}[theorem]{Lemma}

\newcommand{\PreserveBackslash}[1]{\let\temp=\\#1\let\\=\temp}
\newcolumntype{C}[1]{>{\PreserveBackslash\centering}p{#1}} 
\newcolumntype{R}[1]{>{\PreserveBackslash\raggedleft}p{#1}} 
\newcolumntype{L}[1]{>{\PreserveBackslash\raggedright}p{#1}} 

\newcolumntype{Y}{>{\centering\arraybackslash}b{0.7cm}}
\newcolumntype{M}{>{\centering\arraybackslash}b{2.1cm}}
\newcolumntype{Z}{>{\centering\arraybackslash}b{1.75cm}}
\newcolumntype{G}{>{\centering\arraybackslash}m{1.5cm}}
\newcolumntype{Q}{>{\centering\arraybackslash}b{3.75cm}}
\newcolumntype{A}{>{\centering\arraybackslash}m{1.5cm}}
\newcolumntype{B}{>{\centering\arraybackslash}m{0.6667cm}}
\newcolumntype{T}{>{\centering\arraybackslash}b{6.5ex}}
\newcolumntype{S}{>{\centering\arraybackslash}b{13.25ex}}
\newcolumntype{?}{!{\vrule width 1pt}}
\newcolumntype{+}{!{\vrule width 2pt}}

\DeclareFontFamily{U}{mathx}{}
\DeclareFontShape{U}{mathx}{m}{n}{<-> mathx10}{}
\DeclareSymbolFont{mathx}{U}{mathx}{m}{n}
\DeclareMathAccent{\widecheck}{0}{mathx}{"71}

\definecolor{bluee}{RGB}{0, 82, 200}
\newcommand{\mbf}[1]{\boldsymbol{\mathrm{#1}}}
\newcommand{\norm}[1]{\left\lVert#1\right\rVert}

\newcommand{\psp}{\hspace{0.1em}}
\newcommand{\pspp}{\hspace{0.05em}}
\newcommand{\nsp}{\hspace{-0.1em}}
\newcommand{\nspp}{\hspace{-0.05em}}

\newcommand{\diag}{\operatorname{diag}}

\newcommand{\ree}[1]{\text{Re}\nsp\left\{#1\right\}}
\newcommand{\imm}[1]{\text{Im}\nsp\left\{#1\right\}}

\newcommand{\herm}[1]{{#1}^{\texttt{H}}}

\title{Source Enumeration using the Distribution of Angles: A Robust and Parameter-Free Approach}

\author{
	\IEEEauthorblockN{
		Gokularam Muthukrishnan\textsuperscript{*}, Siva Shanmugam\textsuperscript{*}, Sheetal Kalyani\textsuperscript{*}
		\thanks{\noindent\textsuperscript{*}The authors are with the Department of Electrical Engineering, Indian Institute of Technology Madras, Chennai 600036, India (e-mail: \{ee17d400@smail, ee23d001@smail, skalyani@ee\}.iitm.ac.in).
        }}
}

\begin{document}

\maketitle

\begin{abstract}
Source enumeration, the task of estimating the number of sources from the signal received by the array of antennas, is a critical problem in array signal processing. Numerous methods have been proposed to estimate the number of sources under white or colored Gaussian noise. However, their performance degrades significantly in the presence of a limited number of observations and/or a large number of sources. In this work, we propose a method leveraging the distribution of angles that performs well in (a) independent Gaussian, (b) spatially colored Gaussian, and (c) heavy-tailed noise, even when the number of sources is large.  We support the supremacy of our algorithm over state-of-the-art methods with extensive simulation results.
\end{abstract}

\begin{IEEEkeywords}
Source Enumeration, Distribution of Angles, Model Order Selection, Array Signal Processing.
\end{IEEEkeywords}

\section{Introduction} \label{sec:intro}
 
\IEEEPARstart{S}{ource} enumeration, which essentially estimates the rank of the matrix comprising multiple noisy observations (termed `\textit{snapshots}') of the signal received by an array of antennas, appears in many array signal processing algorithms such as multiple-input multiple-output (MIMO) wireless communications \cite{yuan_icc_2017, ke_tsp_2021} and parametric direction-of-arrival (DOA) estimation methods like MUSIC \cite{MUSIC_1986} and ESPRIT \cite{ESPRIT_1986}. However, it remains a challenging task due to various factors and continues to garner the researchers' interest to date.

Several methods have been proposed for this task. In \cite{kritchman2009non}, a random matrix theory-based scheme utilizing the sample covariance matrix has been proposed assuming spatially white Gaussian noise, and the shrinkage estimator of noise covariance has been harnessed in \cite{huang_lsmdl_2013} and \cite{huang_2015}. However, these methods perform poorly under colored noise. Signal subspace matching (SSM) \cite{ssm_2021} relies on the affinity between the observed and hypothesized signal spaces, and the improvised variant in \cite{issm_matiwax_2022} leverages the rotational invariance property of the array steering matrix. Although both can handle colored noise, they are ineffective when the transmit signals are of unequal powers. A recent work \cite{zhang_adlls_2024} uses the coefficient of shrinkage estimators like \cite{huang_lsmdl_2013} and \cite{huang_2015} and can also handle colored noise, but it fails when the number of sources is large. 
 
Though numerous methods exist, with some equipped to handle colored noise, their utility is limited due to the restrictive underlying assumptions (e.g., the small rank assumption in \cite{zhang_adlls_2024}) and/or the presence of the hyperparameters that need to be tuned or set based on \textit{a priori} knowledge (e.g., $\delta$ in \cite{ssm_2021}). Also, when  the noise is heavy-tailed, many techniques perform poorly. Moreover, in emerging applications such as integrated sensing and communication (ISAC), the number of targets is typically large \cite{zhang2024integrated, liu2020joint}, and existing schemes like \cite{issm_matiwax_2022}, \cite{zhang_adlls_2024}, and \cite{ke_tsp_2021} fail to cater to this need (see Sec. \ref{sec:simulation} for more details). Additionally, in applications like automotive radar, the number of available snapshots is limited due to non-stationarity, often less than the number of antenna elements \cite{sun2020mimo}. Thus, there is a need for an efficient approach that performs satisfactorily in a plethora of challenging settings. 

We propose a generalized approach for rank estimation that performs substantially better than existing source enumeration schemes in a wide range of scenarios and is hyperparameter-free. It is based on the statistical distribution of angles between high-dimensional points \cite{cai2013distributions}, which have been adopted in other signal processing tasks \cite{menon2019structured,vishnu_subspace_tsp_2020}. Our technique, SEDAN (\textit{source enumeration using the distribution of angles}), can handle  colored noise as well as heavy-tailed noise and  exhibits superior performance when the number of sources is large. Ours is the first work to leverage the statistical distribution of angles for source enumeration, and it exhibits superior performance compared to state-of-the-art techniques in several settings. While the existing methods just consider the eigenvalues of the empirical covariance matrix of the measurements, the proposed scheme exploits additional information about the signal and noise subspaces through the statistical distribution of angles, which renders a more detailed description of the subspaces than the set of singular values.
\vspace{0.85ex}

\noindent\textbf{Basic Notations:} 
We use the shorthand $a\wedge b$ for $\min(a,b)$, $a$, $b\in\mathbb{R}$. $\ree{\cdot}$ and $\imm{\cdot}$ provide the real and imaginary parts. We use boldface uppercase (lowercase) letters for matrices (vectors).  $\norm{\cdot}_p^{}$ and $\norm{\cdot}_{\texttt{F}}^{}$ denote the $\ell_p^{}$ and Frobenius norms. $\mathscr{R}(\mbf{B})$, $\mbf{B}^\top_{}$, and $\herm{\mbf{B}}_{}$ indicate the range space, the transpose and adjoint of $\mbf{B}$. $\mbf{I}_M^{}$ denotes the identity matrix of dimension $M$.  The diagonal matrix with entries $b_1^{},\, \ldots,\, b_M^{}$ is written as $\diag([b_1^{},\, \ldots,\, b_M^{}])$. The real and circularly symmetric complex Gaussian distributions with mean $\mbf{\mu}$ and covariance $\mbf{C}$ are denoted by $\mathcal{N}(\mbf{\mu},\mbf{C})$  and $\mathcal{CN}(\mbf{\mu},\mbf{C})$, respectively. 

\section{Background} \label{sec:problem}

Consider an $M$-element uniform linear array (ULA) impinged upon by the narrowband signals from $r < M$ sources at the angles $\{\phi_i^{}\}_{i=1}^r$. The received signal at instance $n$, $\mbf{x}(n)\in\mathbb{C}^{M\times 1}_{}$, is given by $\mbf{x}(n) = \mbf{A} \mbf{s}(n) +\mbf{z}(n)$, where $\mbf{s}(n)\in\mathbb{C}^{r\times 1}_{}$ is the collection of transmit signals from $r$ sources at instance $n$, $\mbf{A}\in\mathbb{C}^{M\times r}_{}$ is the steering matrix with the array steering vectors $\{\mbf{a}(\phi_i^{})\}_{i=1}^r$  as columns and $\mbf{z}(n)\in \mathbb{C}^{M\times 1}_{}$ is the zero-mean additive noise with covariance $\mbf{C}_{\mbf z}^{}$. We consider non-coherent Gaussian transmit signals, $\mbf{s}(n) \sim \mathcal{CN}(\mbf{0}, \mbf{C}_{\mbf s}^{})$, where $\mbf{C}_{\mbf{s}}^{} = \diag\big(\big[\sigma_{s_1^{}}^2,\, \ldots,\,\sigma_{s_r^{}}^2\big]\big)$. Hence, the covariance of the observations is $\mbf{C}_{\mbf x}^{}=\mbf{C}_0^{}+\mbf{C}_{\mbf z}^{}$,  where $\mbf{C}_0^{}=\mbf{A}\mbf{C}_{\mbf s}^{}\herm{\mbf{A}}_{}$ is a positive definite matrix of rank $r$. Given the spectral decomposition $\mbf{C}_{\mbf x}^{}=\mbf{U}\mbf{\Lambda}\herm{\mbf{U}}_{}$, let $\mbf{U}=[\mbf{U}_k^{} \ \mbf{U}_k^{\perp}]$ be  the partition of $\mbf{U}$ into its first $k$ columns (corresponding to $k$ largest eigenvalues) and remaining $M-k$ columns, $k=0,\,\ldots,\,M$. Let $\mbf{X}=[\,\mbf{x}(1)\,\cdots\,\mbf{x}(N)\,]\in\mathbb{C}^{M\times N}_{}$ be the collection of $N>r$ independent observations of the received signal; hence, $\mbf{X} = \mbf{L} + \mbf{Z}$, where $\mbf{L}=\mbf{A}\mbf{S}$ denote the collection of noise-free signals received by the array, and $\mbf{S}\in\mathbb{C}^{r\times N}_{}$ and $\mbf{Z}\in\mathbb{C}^{M\times N}_{}$ are respectively the accumulations of $\{\mbf{s}(n)\}_{n=1}^N$ and $\{\mbf{z}(n)\}_{n=1}^N$. Thus, source enumeration is the problem of estimating the rank $r$ of $\mbf{L}$ from the noisy observations $\mbf{X}$.

In this work, we leverage the statistical distribution of angles subtended by the points for the task of source enumeration. The distribution of angles between two vectors has been studied in detail in \cite{cai2013distributions}, and it can be approximated by a Gaussian distribution in higher dimensions:
\begin{lemma}[{\hspace{0.01ex}\cite{cai2013distributions,vishnu_subspace_tsp_2020}}] 
\label{lemma:gaussian}
        Let $\mbf{y}_i^{}$, $\mbf{y}_j^{} \in \mathbb R^d_{}$ be two independent random vectors sampled uniformly at random from the unit hypersphere $\mathbb S^{d-1}_{}=\{\mbf{y} \in \mathbb R^d \,|\, \lVert \mbf{y} \rVert_2^{} = 1 \}$. The principal angle between these vectors, i.e., $ \arccos(\mbf{y}_i^\top \mbf{y}_j^{}) \in [0, \pi]$, converges weakly in distribution to $\mathcal{N}\big(\frac{\pi}{2}, \frac{1}{d-2}\big)$ as $d \to \infty$.
\end{lemma}
\noindent The Gaussian approximation holds good for the dimensions $d\geq 5$ \cite{cai2013distributions}. We use this result to design our source enumeration scheme in the sequel. Bhattacharyya distance (BD) \cite{bhattacharyya1943measure} is a popular metric for measuring \textit{distances} between probability distributions\footnote{We consider Bhattacharyya distance since it is a symmetric measure, but other divergences could be used.}. BD between $\mathcal{N}(\mu_1^{}, \sigma_1^2)$ and $ \mathcal{N}(\mu_2^{}, \sigma_2^2)$ is
\begin{equation*} \label{eq:d_bhatt}
    D_{\mathcal{N}}^{(\texttt{B})}(\mu_1^{},\,\sigma_1^2,\,\mu_2^{},\,\sigma_2^2) = \frac{1}{4} \frac{(\mu_{1}^{} - \mu_2^{})^2_{}}{\sigma_{1}^2 + \sigma_2^2} + \frac{1}{2}\log\bigg(\frac{\sigma_{1}^2 + \sigma_2^2}{2\, \sigma_{1}^{} \sigma_2^{}}\bigg) .
\end{equation*}
We use its empirical variant \cite{jain1976estimate} in this work. Let $\mathcal{S}_1^{}$ and $\mathcal{S}_2^{}$ be finite sets with real entries, and $\widehat{\mu}_i^{}$ and $\widehat{\sigma}^2_i$ be the sample mean and sample variance of the elements of $\mathcal{S}_i^{}$. We denote
\begin{equation} \label{eq:d_bhatt_emp}
\rho\,(\mathcal{S}_1^{},\mathcal{S}_2^{}) = D_{\mathcal{N}}^{(\texttt{B})}(\widehat{\mu}_1^{},\,\widehat{\sigma}_1^2,\,\widehat{\mu}_2^{},\,\widehat{\sigma}_2^2).
\end{equation}
For matrix $\mbf{B}=[\,\mbf{b}(1)\,\cdots\,\mbf{b}(N)\,]\in\mathbb{C}^{L\times N}_{}$, we define the set 
\begin{equation} \label{eq:ang_set}
    \mathcal{K}({\mbf{B}}) \nsp = \nsp \Big\{ \!    \arccos\nsp\Big( \ree{\nspp\tfrac{\herm{{\mbf{b}}(i)}_{} {\mbf{b}}(j)}{\norm{{\mbf{b}}(i)}_2^{}\norm{{\mbf{b}}(j)}_2^{}}\nsp}\nsp\!\Big)   \Big|\psp   1\nspp\leq\nspp i\nspp <\nspp j \nspp\leq\nspp N\Big\} .\!\nspp
\end{equation} 

\section{Proposed Method} \label{sec:proposed}

We motivate our method by assuming the availability of the observations' covariance\footnote{However, we consider the noise covariance $\mbf{C}_{\mbf z}^{}$ to be unknown as it cannot be estimated without knowing the number of sources: When $\mbf{C}_{\mbf z}^{}$ is known, the number of sources is the same as the number of generalized eigenvalues of the pair $(\mbf{C}_{\mbf x}^{},\mbf{C}_{\mbf z}^{})$ greater than 1\cite[Sec. 7.3.2]{johnson1992array}.} $\mbf{C}_{\mbf x}^{}$. Practically, this can be substituted with its estimate, a standard approach in array signal processing  \cite{MUSIC_1986, ESPRIT_1986, li2003robust}. We also assume that the signal subspace can be identified from the $r$ dominant eigenvectors of $\mbf{C}_{\mbf x}^{}$, i.e., $\mathscr{R}(\mbf{L})=\mathscr{R}(\mbf{C}_0^{})=\mathscr{R}(\mbf{U}_r^{})$ \cite{johnson1992array}; this is crucial for any source enumeration scheme because  severe \textit{mixing} of the signal and noise subspaces would otherwise render the sources indistinguishable from noise. Let the hypothesized number of sources be $k$; we define an estimate for the signal component from the observations $\mbf{X}$ through the least squares principle as
\begin{equation}\label{eq:sig_comp_est}
    \widehat{\mbf{L}}_k^{} = \underset{\mathscr{R}(\mbf{G}) \,=\, \mathscr{R}(\mbf{U}_k^{})}{\operatorname{argmin}}\,\norm{\mbf{X}-\mbf{G}}_{\texttt{F}}^{2} = \mbf{U} k^{}\herm{\mbf{U}}_k\,\mbf{X}
\end{equation} 
and the corresponding noise component estimate as $\widehat{\mbf{Z}}_k^{} = \mbf{X} - \widehat{\mbf{L}}_k^{} = \mbf{U}_k^{\perp}\,\herm{(\mbf{U}_k^{\perp})}_{}\,\mbf{X}$. Suppose the \textit{directions} $\frac{{\mbf{z}}(n)}{\norm{{\mbf{z}}(n)}_2^{}}$, $n=1,\,\ldots,\,N$, are distributed uniformly over the hypersphere\footnote{This holds when the distribution of $\mbf{z}(n)$ is circularly symmetric with a density function $f$ that is isotropic in the $\ell_2^{}$-norm, i.e., $f(\mbf{z})$ depends on $\mbf{z}$ only through $\norm{\mbf{z}}_2^{}$; note that the coordinates of $\mbf{z}(n)$ need not be independent.} $\mathbb{S}^{M-1}_{}$, from Lemma \ref{lemma:gaussian}, we know that the distribution of angles subtended by the independent samples\footnote{Note that the angles that involve the same point are only pairwise independent, even if the points themselves are independently distributed \cite{cai2012phase}.} $\mbf{z}(i)$ and $\mbf{z}(j)$ can be approximated well by $\mathcal{N}\big(\frac{\pi}{2}, \frac{1}{2(M-1)}\big)$, especially when $M \geq 3\pspp$, and $\mathcal{K}({\mbf{Z}})$ comprises of $\frac{N(N-1)}{2}$ observations from this distribution.

Furthermore, when $k\geq r$, $\widehat{\mbf{Z}}_k^{} = \mbf{U}_k^{\perp}\,\herm{(\mbf{U}_k^{\perp})}_{}\,\mbf{Z}$. Consequently, the  pairwise angles between the columns of $\widehat{\mbf{Z}}_k^{}$ can be expressed as the angles between the corresponding columns in ${\mbf{V}}_k^{} = \herm{(\mbf{U}_k^{\perp})}_{}\,\mbf{Z}\pspp$: Since the columns of ${\mbf{V}}_k^{}$ are linearly (and statistically) independent with the directions distributed uniformly over $\mathbb{S}^{M-k- 1}_{}$, the elements of  $\mathcal{K}(\widehat{\mbf{Z}}_k^{})$ are the samples from $\mathcal{N}\big(\frac{\pi}{2},\frac{1}{2(M-k-1)}\big)$. Hence, we can estimate the rank as the smallest $k$ that results in an \textit{empirical}  distribution of the angles in $\mathcal{K}\big(\widehat{\mbf{Z}}_k^{}\big)$ which is closest to this distribution: $   \widehat{r}=\min \big\{ k\,|\,D_{\mathcal{N}}^{(\texttt{B})} \big(\widehat{\mu}_k^{},\,\widehat{\sigma}_k^2,\,\tfrac{\pi}{2},\,\tfrac{1}{2(M-k-1)}\big) \leq \tau \big\} $, where $\widehat{\mu}_k^{}$ and $\widehat{\sigma}^2_k$ are the sample mean and variance of the elements of $\mathcal{K}\big(\widehat{\mbf{Z}}_k^{}\big)$ (using independent samples if needed \cite[Lemma 3]{vishnu_subspace_tsp_2020}); $\tau$  is a high probability bound on the empirical Bhattacharyya distance based on its mean and variance \cite{jain1976estimate}.

\begin{figure*}[h]
\begin{subfigure}{.5\textwidth}
\centering
\includegraphics[width=.865\linewidth]{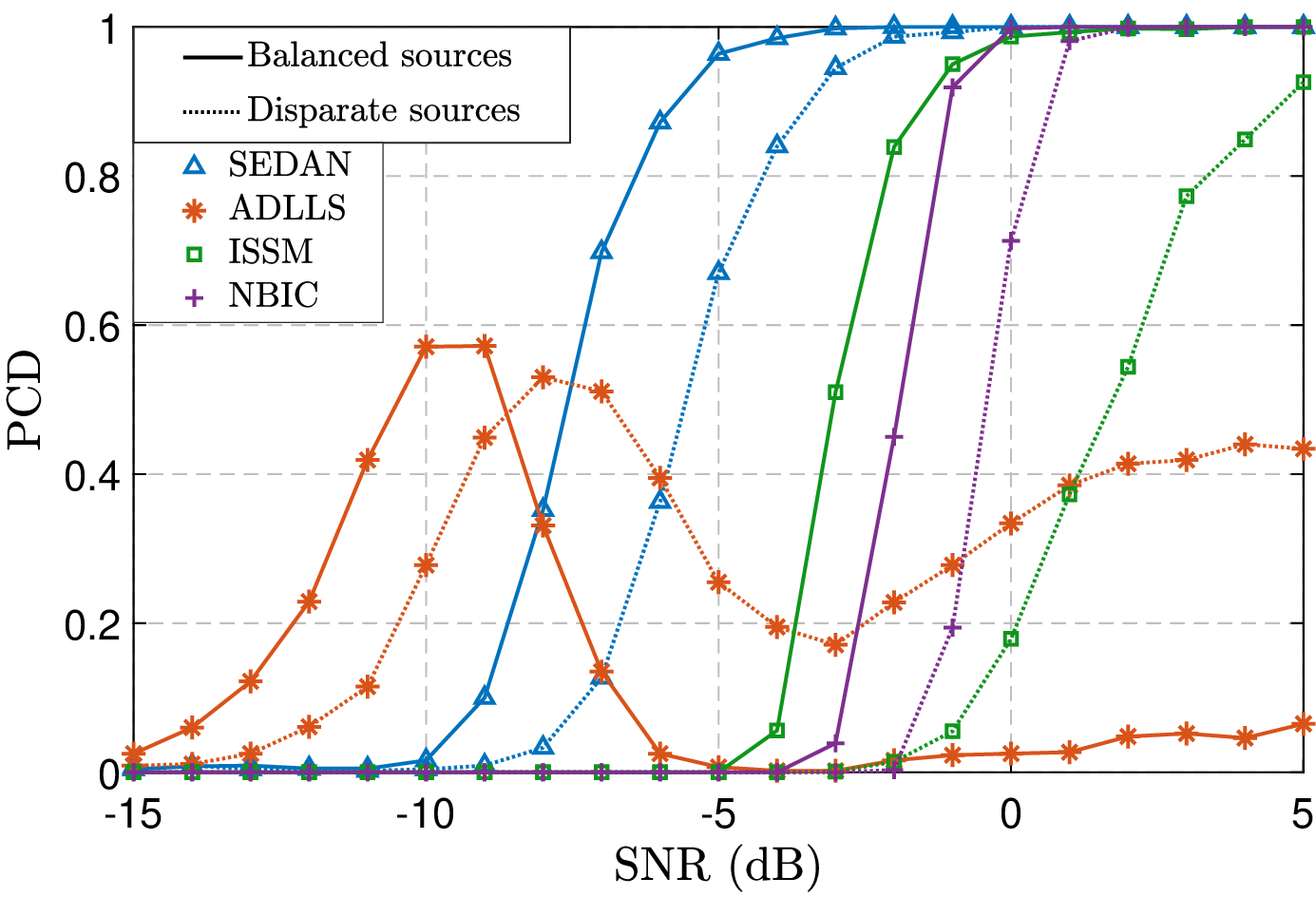}
\caption*{\vspace{-1em}}
\end{subfigure}
\begin{subfigure}{.5\textwidth}
\centering
\includegraphics[width=.865\linewidth]{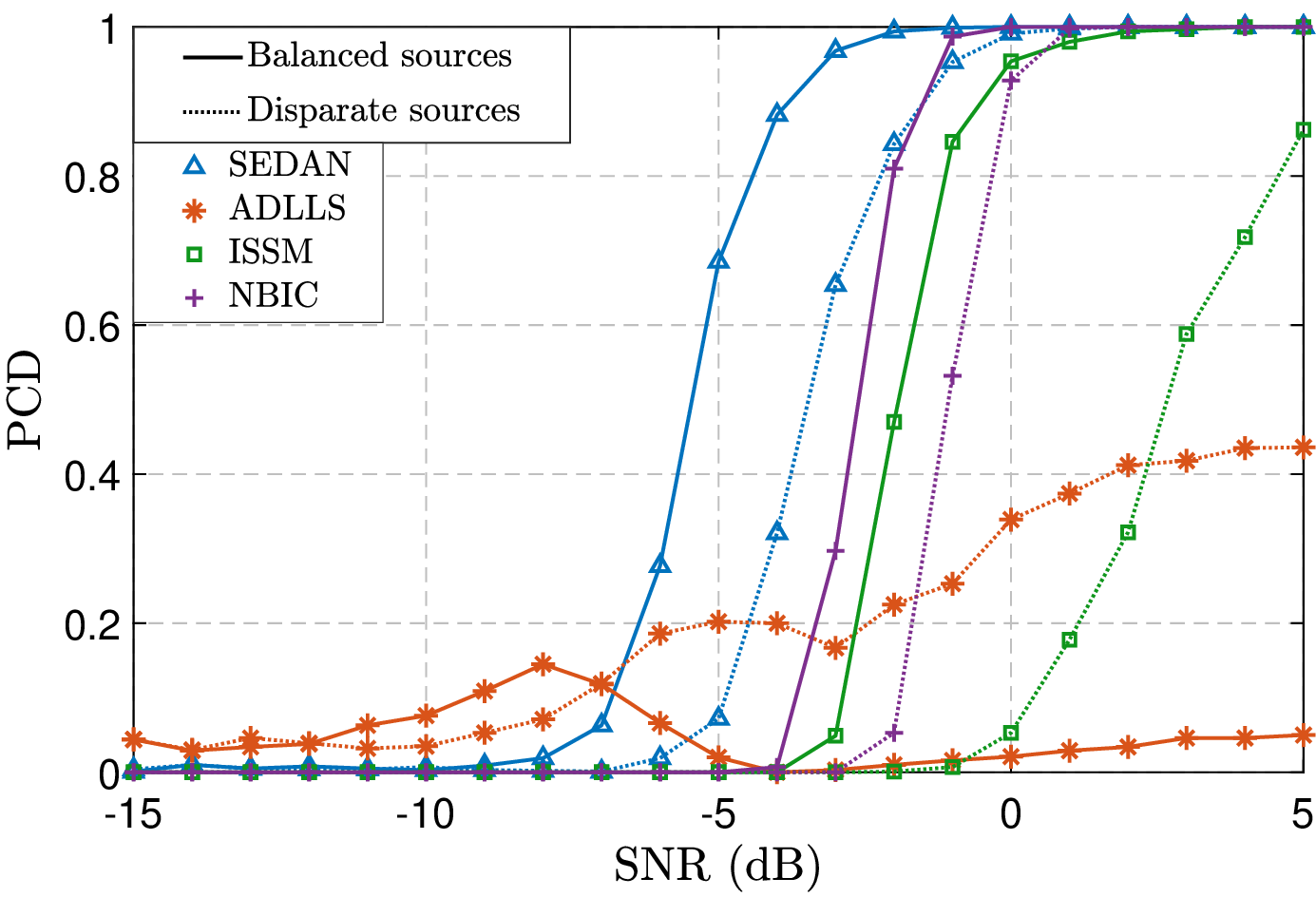}
\caption*{\vspace{-1em}}
\end{subfigure}
\begin{subfigure}{.5\textwidth}
\centering
\includegraphics[width=.865\linewidth]{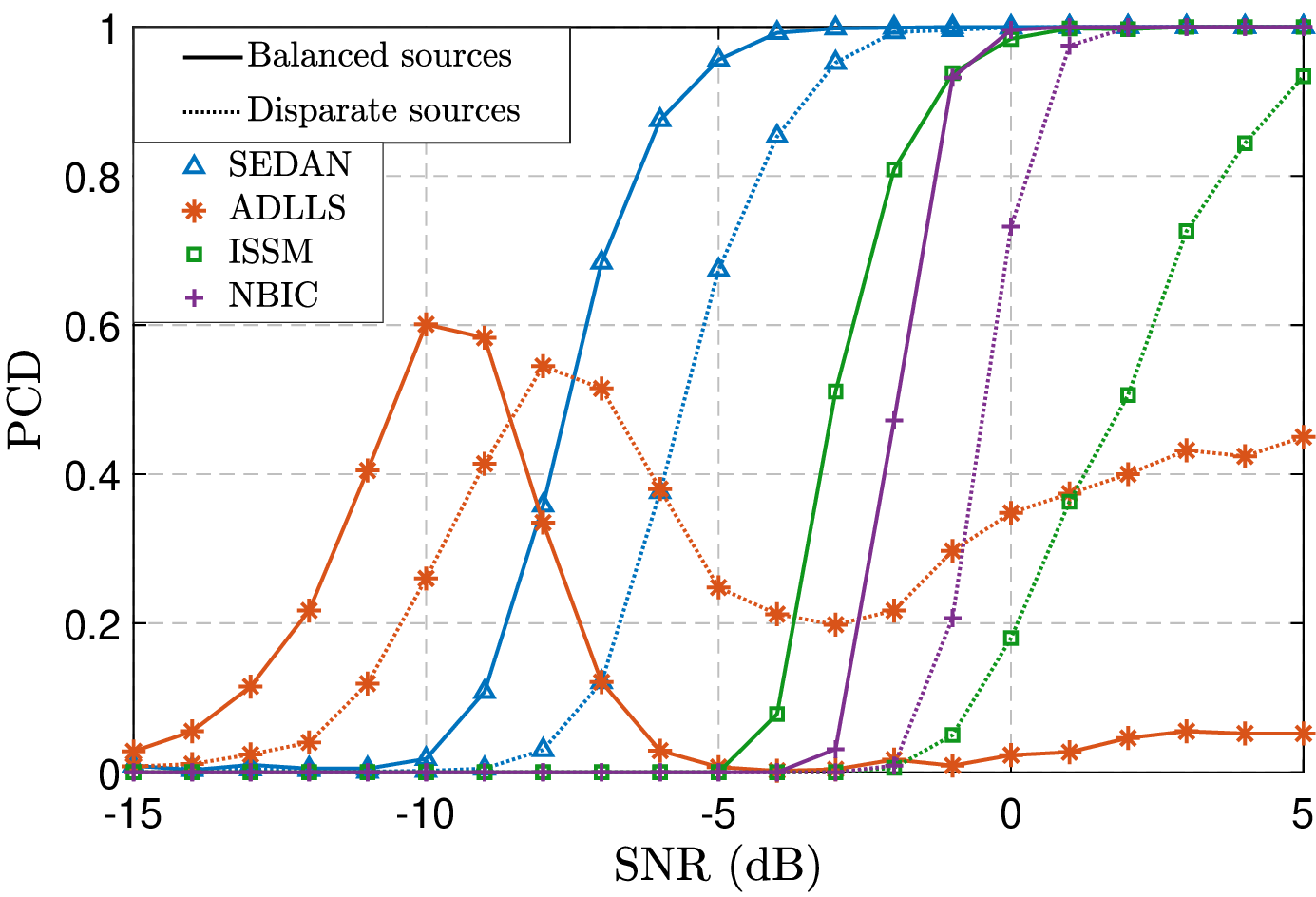}
\end{subfigure}
\begin{subfigure}{.5\textwidth}
\centering
\includegraphics[width=.865\linewidth]{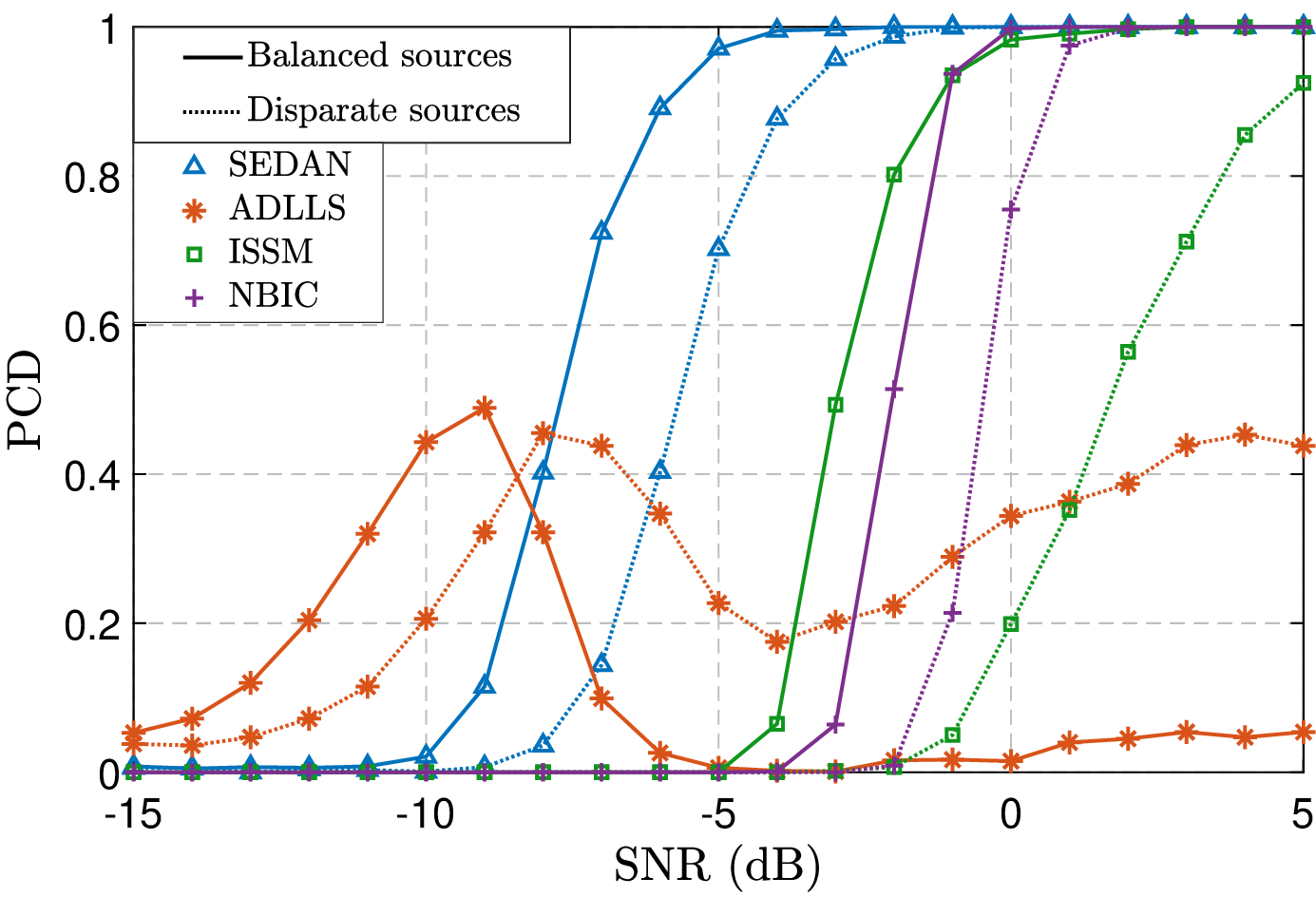}
\end{subfigure}
\caption{PCD for Case 1 under i.i.d. Gaussian noise (top-left), BCN (top-right), $\ell_2^{}$-noise (bottom-left), and GMN (bottom-right).}
\label{fig:case1_results}
\end{figure*}

The isotropic assumption on the noise density gives us a framework to work with, but it is still restrictive. For instance, when the coordinates of the noise vector are of different power, even if they are independent, the noise density will not be isotropic. Hence, we generalize this model: We assume that the angles between the noise vectors are approximately Gaussian-distributed with \textit{arbitrary} and unknown mean and variance\footnote{This assumption is empirically validated in Appendix \ref{appn:emp_dist_angles}.}. This generalized model has been used in \cite{vishnu_subspace_tsp_2020}. Since we do not know the parameters of such a reference distribution, we resort to an approach different from the one we motivated above. For the hypothesized rank $k$, we define a score 
\begin{equation}\label{eq:eta_k}
    \eta_k^{}=\rho\big(\mathcal{K}\big(\widehat{\mbf{Z}}_{k}^{}\big)^{},\mathcal{K}\big(\widehat{\mbf{Z}}_{k-1}^{}\big)\big) .
\end{equation}
As $k$ increases, the behavior of $\eta_k^{}$ will be as follows. 
\begin{itemize}
    \item For $k<r+1$, $\widehat{\mbf{Z}}_{k-1}^{}$ will have at least one signal component, and $\widehat{\mbf{Z}}_{k}^{}$ will have one signal component less. Hence, the distributions of angles in $\mathcal{K}\big(\widehat{\mbf{Z}}_{k}^{}\big)$ and $\mathcal{K}\big(\widehat{\mbf{Z}}_{k-1}^{}\big)$ will be distinct, and $\eta_k^{}$ will be away from zero.
    \item When $k=r+1\pspp$, both $\widehat{\mbf{Z}}_{k}^{}$ and $\widehat{\mbf{Z}}_{k-1}^{}$ will have only the noise components. Thus, $\eta_k^{}$ will be close to zero.
    \item For $k>r+1$, $\eta_k^{}$ will change less drastically, and slicing a subspace out of the residue would affect the noise less, as it would not be concentrated in any subspace.
\end{itemize}

Thus, the trend in $\eta_k^{}$ is indicative of the rank. We estimate the rank as hypothesis $k$, after which $\eta_{k}^{}$ drops drastically: 
\begin{equation} \label{eqn:mintau}
    \widehat r  =  \underset{k\in \{1,\,\ldots,\,M-2\}}{\arg\max} \quad \eta_{k}^{} -\eta_{k+1}^{}\pspp.
\end{equation} 
In practice, $\mbf{C}_{\mbf x}^{}$ is unknown and has to be estimated from $\mbf{X}$. Any estimate can be used \cite{daniels2001shrinkage,ledoit2003honey,romanov2023tyler}, but for simplicity, we use sample covariance. The scheme is presented in Algorithm \ref{algo:SEDAN}.

\begin{algorithm}[h!]
\caption{Source Enumeration using the Distribution of Angles (SEDAN).}
\label{algo:SEDAN}
\begin{algorithmic}[1]
\State {\bfseries Input:} Observations of received signal $\mbf{X}\in\mathbb{C}^{M\times N}_{}$.
\vspace{0.5ex}
\State Obtain $\mbf{U}$ from the spectral decomposition of $\tfrac{1}{N}\mbf{X}\herm{\mbf{X}}_{}$.
\vspace{0.5ex}
\State Initialize $\widehat{\mbf{Z}}_0^{}=\mbf{X}$ and construct $\mathcal{K}\big(\widehat{\mbf{Z}}_{0}^{}\big)$.
\vspace{0.5ex}
\For{$k=1,\,\ldots,\,M\wedge N-1$}
\vspace{0.5ex}
\State Obtain $\widehat{\mbf{Z}}_k^{} = \mbf{U}_k^{\perp}\,\herm{(\mbf{U}_k^{\perp})}_{}\,\mbf{X}$ and $\mathcal{K}\big(\widehat{\mbf{Z}}_{k}^{}\big)$ as in \eqref{eq:ang_set}.
\vspace{0.5ex}
\State Compute the score $\eta_k^{}$  as in \eqref{eq:eta_k}.
\vspace{0.5ex}
\vspace{0.5ex}
\EndFor
\vspace{0.5ex}
\State {\bfseries Output:} Estimated number of sources $\widehat r $ from \eqref{eqn:mintau}.
\end{algorithmic}
\end{algorithm}

\begin{figure*}[h]
\begin{subfigure}{.5\textwidth}
\centering
\includegraphics[width=.865\linewidth]{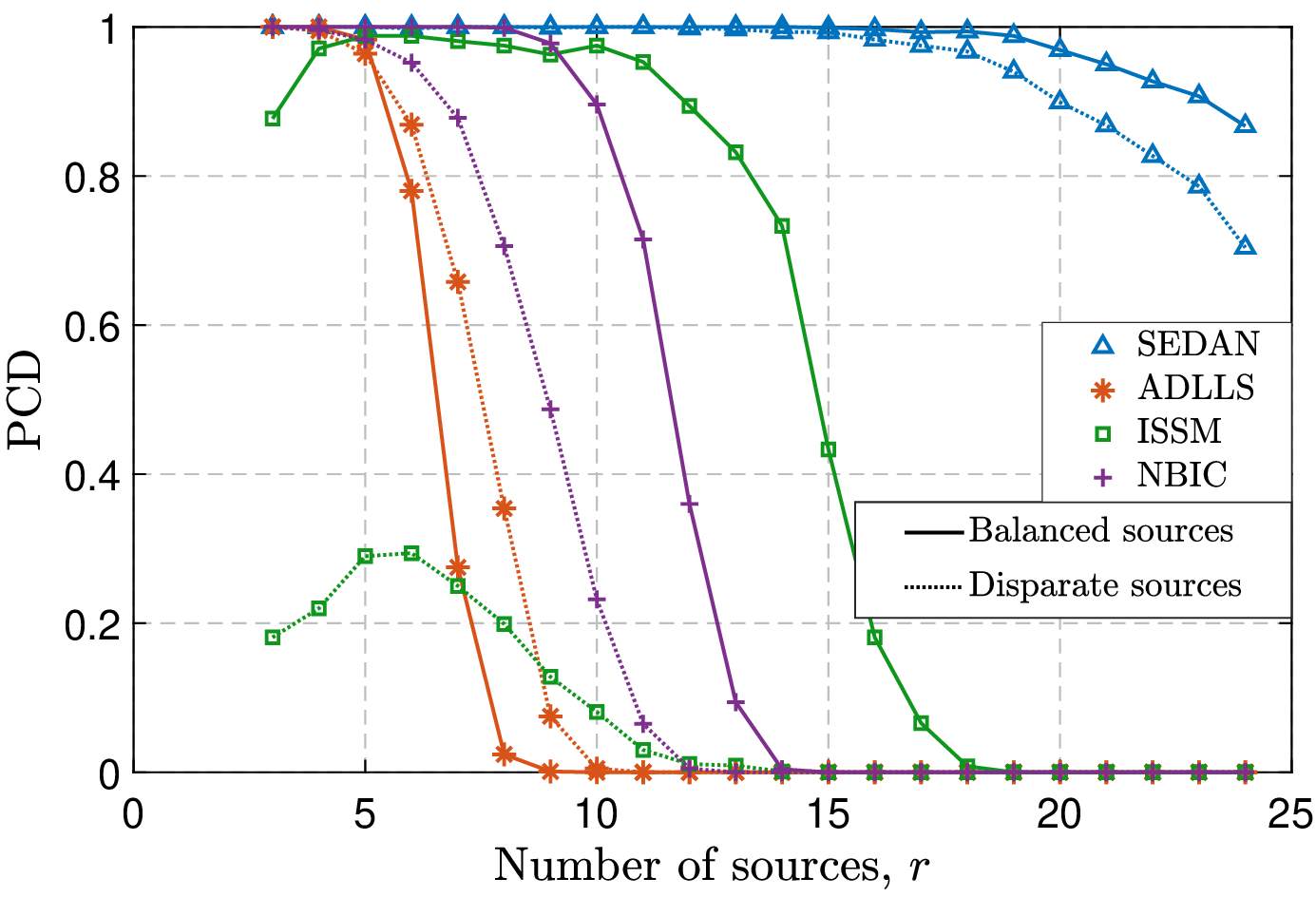}
\caption*{\vspace{-1em}}
\end{subfigure}
\begin{subfigure}{.5\textwidth}
\centering
\includegraphics[width=.865\linewidth]{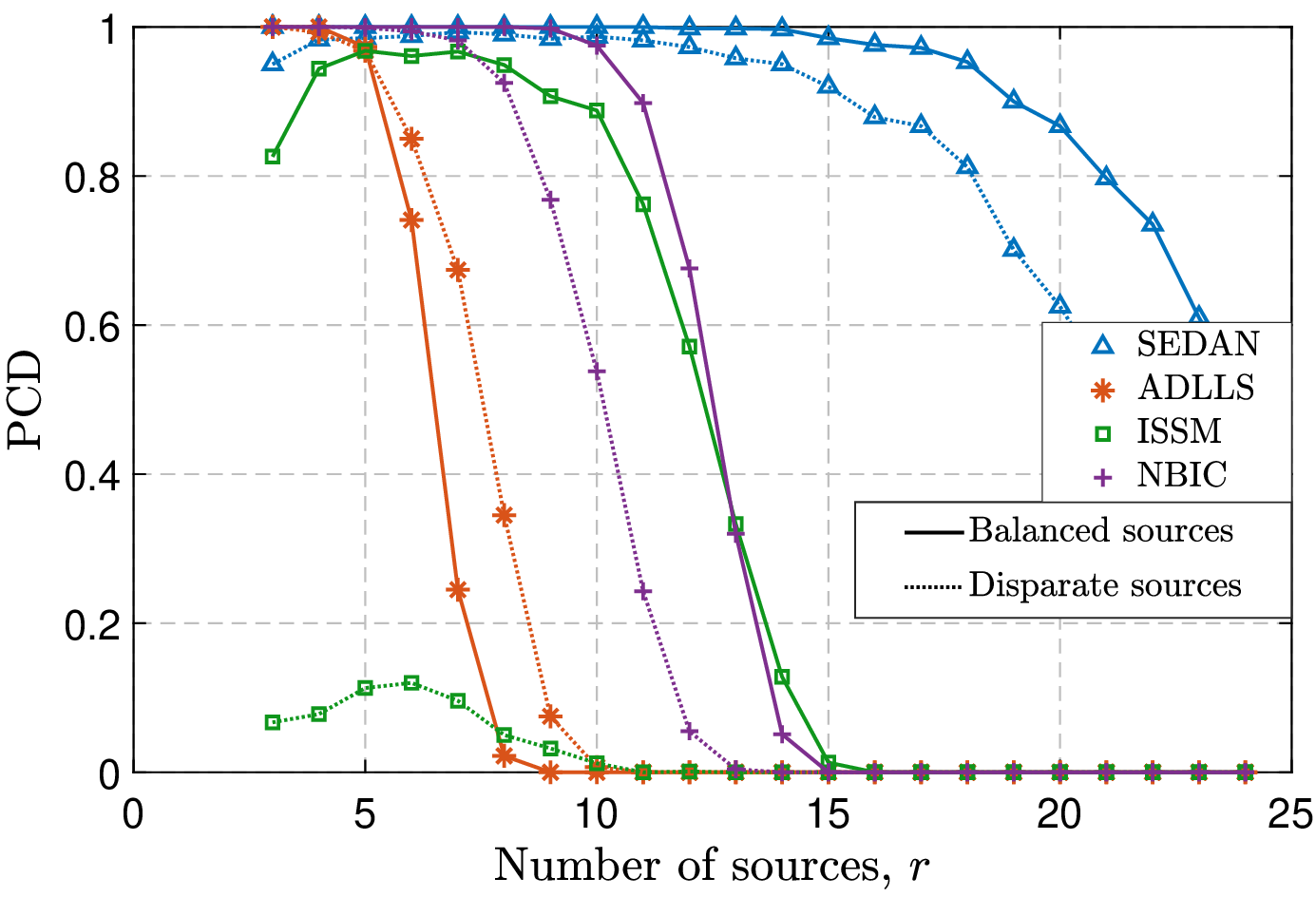}
\caption*{\vspace{-1em}}
\end{subfigure}
\begin{subfigure}{.5\textwidth}
\centering
\includegraphics[width=.865\linewidth]{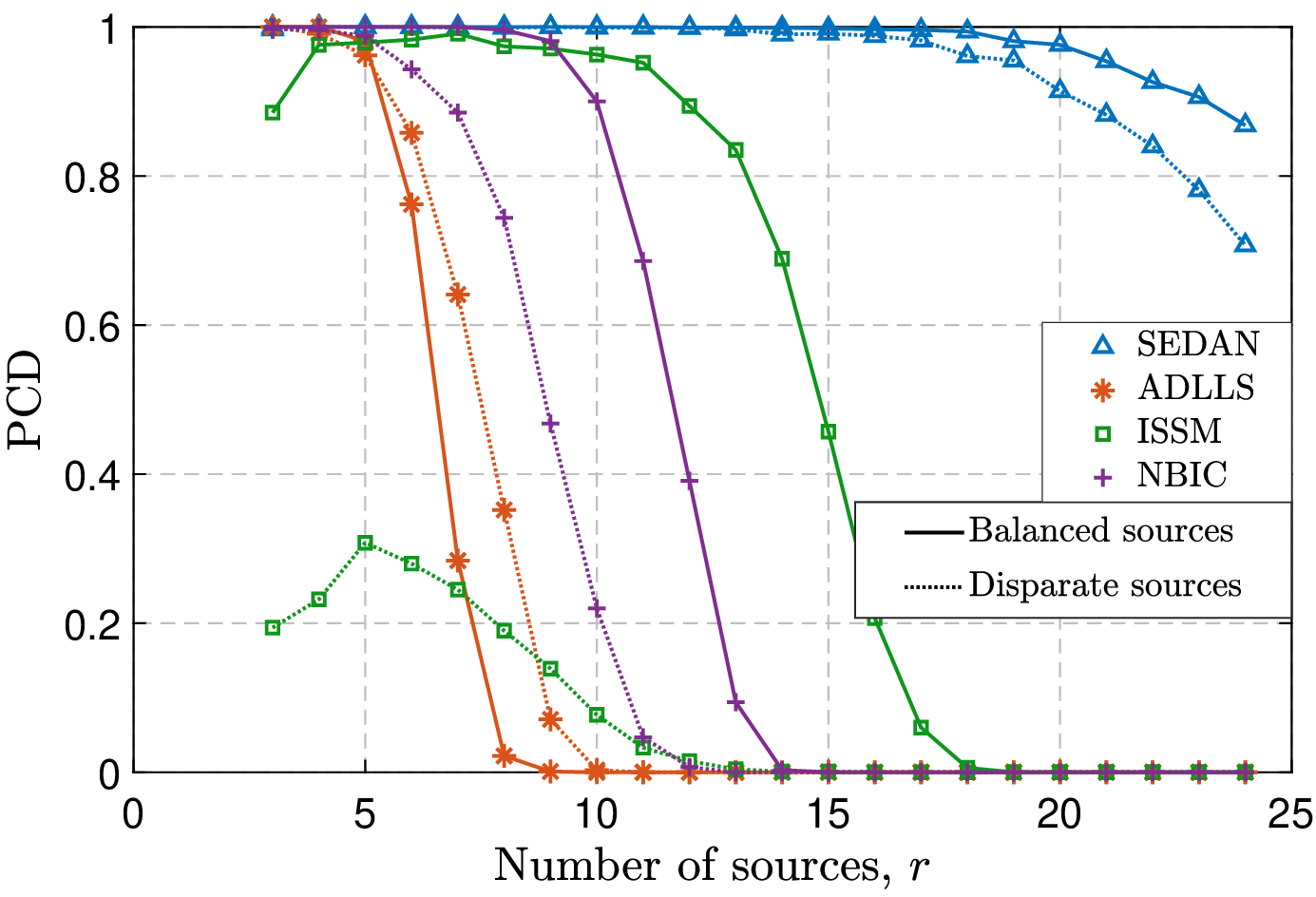}
\end{subfigure}
\begin{subfigure}{.5\textwidth}
\centering
\includegraphics[width=.865\linewidth]{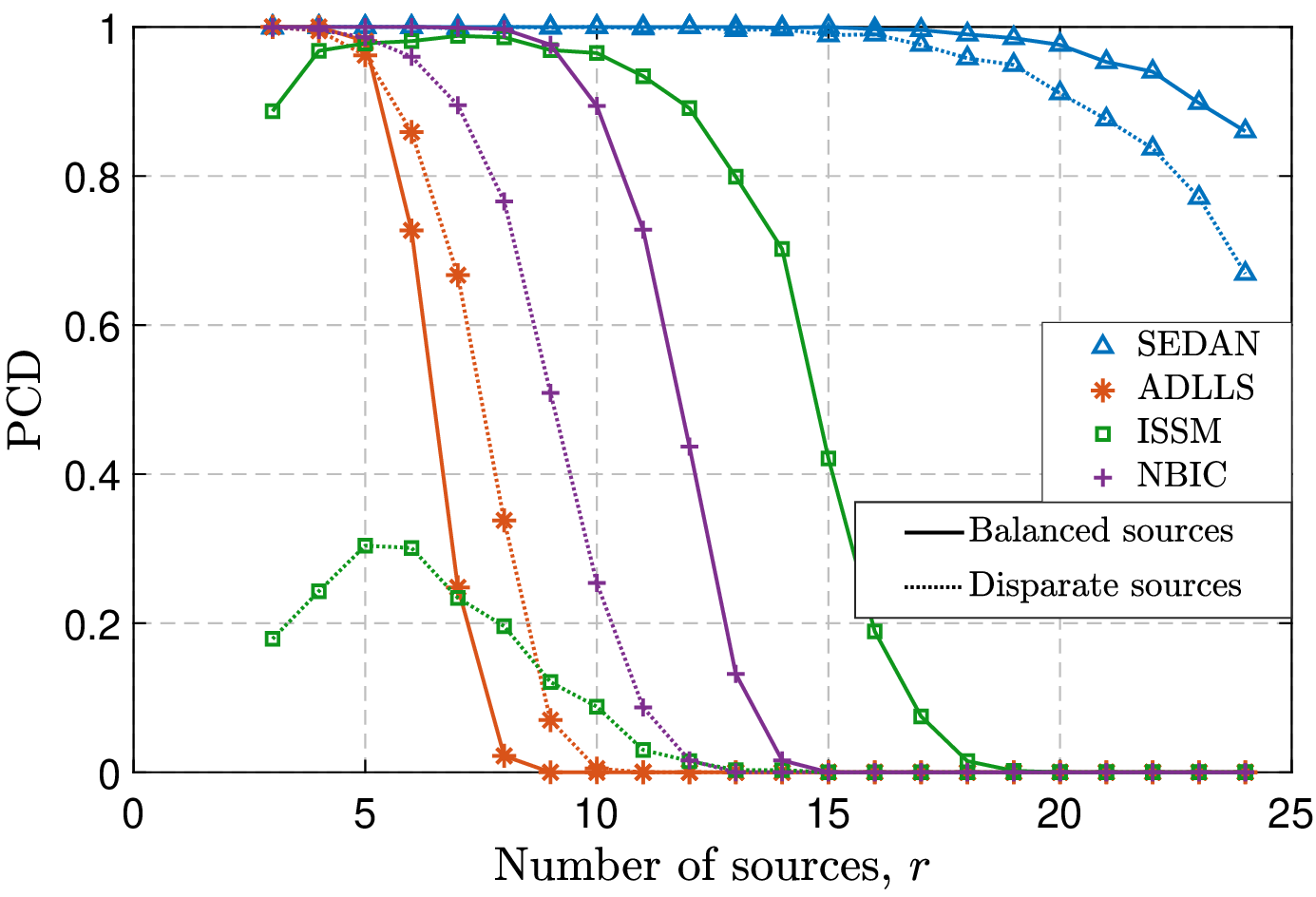}
\end{subfigure}
\caption{PCD for Case 2 under i.i.d. Gaussian noise (top-left), BCN (top-right), $\ell_2^{}$-noise (bottom-left), and GMN (bottom-right).}
\label{fig:case2_results}
\end{figure*}

\setlength{\tabcolsep}{5pt}
\renewcommand{\arraystretch}{0.91}
\begin{table*}[h!]
\caption{PCD for Case 3.}
\label{tab:case3_results}
\begin{center}
\begin{tabular}{? C{1.75em} | C{5em} *{4}{? C{4.25em} | C{4.25em}} ?} 
\clineB{3-10}{2}
\multicolumn{1}{c}{} & \multicolumn{1}{c?}{} &  \multicolumn{2}{c?}{\multirow{1}{*}{\textbf{i.i.d.}}} &  \multicolumn{2}{c?}{\multirow{1}{*}{\textbf{BCN}}} &  \multicolumn{2}{c?}{\multirow{1}{*}{\textbf{$\ell_2^{}$-noise}}} &  \multicolumn{2}{c?}{\multirow{1}{*}{\textbf{GMN}}}
\\[-0.225ex]
\cline{3-10}
\multicolumn{1}{c}{} & \multicolumn{1}{c?}{} & \multirow{1}{*}{${N}\nsp=\nsp40$} & \multirow{1}{*}{${N}\nsp=\nsp80$} & \multirow{1}{*}{${N}\nsp=\nsp40$} & \multirow{1}{*}{${N}\nsp=\nsp80$} & \multirow{1}{*}{${N}\nsp=\nsp40$} & \multirow{1}{*}{${N}\nsp=\nsp80$} & \multirow{1}{*}{${N}\nsp=\nsp40$} & \multirow{1}{*}{${N}\nsp=\nsp80$} 
\\[-0.225ex]
\clineB{3-10}{2}
\specialrule{0.8pt}{2pt}{0pt}
%____________________________________________________________________________________
\multirow{4}{*}{\spheading{\centering%
\begin{tabular}{c}
     \\[-4.85ex]
     \multirow{1}{*}{\textbf{Balanced}}\\[-2.25ex]%
     \multirow{1}{*}{\textbf{sources}$\,$}
\end{tabular}
}}
& \multirow{1}{*}{SEDAN} & \multirow{1}{*}{\textbf{0.998}} & \multirow{1}{*}{\textbf{1}} & \multirow{1}{*}{\textbf{0.971}} & \multirow{1}{*}{\textbf{1}} & \multirow{1}{*}{\textbf{0.998}} & \multirow{1}{*}{\textbf{1}} & \multirow{1}{*}{\textbf{1}} & \multirow{1}{*}{\textbf{1}}
\\[-0.225ex]
\cline{2-10}

& \multirow{1}{*}{ADLLS} & \multirow{1}{*}{0.001} & \multirow{1}{*}{\textbf{1}} & \multirow{1}{*}{0.002} & \multirow{1}{*}{\textbf{1}} & \multirow{1}{*}{0.006} & \multirow{1}{*}{\textbf{1}} & \multirow{1}{*}{0.007} & \multirow{1}{*}{\textbf{1}}
\\[-0.225ex]
\cline{2-10}

& \multirow{1}{*}{ISSM} & \multirow{1}{*}{0.502} & \multirow{1}{*}{0.990} & \multirow{1}{*}{0.047} & \multirow{1}{*}{0.602} & \multirow{1}{*}{0.528} & \multirow{1}{*}{0.987} & \multirow{1}{*}{0.812} & \multirow{1}{*}{0.997}
\\[-0.225ex]
\cline{2-10}

& \multirow{1}{*}{NBIC} & \multirow{1}{*}{0.028} & \multirow{1}{*}{\textbf{1}} & \multirow{1}{*}{0.283} & \multirow{1}{*}{\textbf{1}} & \multirow{1}{*}{0.036} & \multirow{1}{*}{\textbf{1}} & \multirow{1}{*}{0.563} & \multirow{1}{*}{\textbf{1}}
\\[-0.225ex]
\clineB{1-10}{2}
\specialrule{0.8pt}{2pt}{0pt}
%____________________________________________________________________________________
\multirow{4}{*}{\spheading{\centering%
\begin{tabular}{c}
     \\[-4.85ex]
     \multirow{1}{*}{\textbf{Disparate}}\\[-2.25ex]%
     \multirow{1}{*}{\textbf{sources}$\,$}
\end{tabular}
}}
& \multirow{1}{*}{SEDAN} & \multirow{1}{*}{\textbf{0.956}} & \multirow{1}{*}{\textbf{1}} & \multirow{1}{*}{\textbf{0.644}} & \multirow{1}{*}{\textbf{0.959}} & \multirow{1}{*}{\textbf{0.968}} & \multirow{1}{*}{\textbf{0.999}} & \multirow{1}{*}{\textbf{0.987}} & \multirow{1}{*}{\textbf{1}}
\\[-0.225ex]
\cline{2-10}

& \multirow{1}{*}{ADLLS} & \multirow{1}{*}{0.182} & \multirow{1}{*}{\textbf{1}} & \multirow{1}{*}{0.181} & \multirow{1}{*}{\textbf{1}} & \multirow{1}{*}{0.184} & \multirow{1}{*}{\textbf{1}} & \multirow{1}{*}{0.236} & \multirow{1}{*}{\textbf{1}}
\\[-0.225ex]
\cline{2-10}

& \multirow{1}{*}{ISSM} & \multirow{1}{*}{0} & \multirow{1}{*}{0.055} & \multirow{1}{*}{0} & \multirow{1}{*}{0.003} & \multirow{1}{*}{0.002} & \multirow{1}{*}{0.039} & \multirow{1}{*}{0.017} & \multirow{1}{*}{0.129}
\\[-0.225ex]
\cline{2-10}

& \multirow{1}{*}{NBIC} & \multirow{1}{*}{0} & \multirow{1}{*}{0.559} & \multirow{1}{*}{0.002} & \multirow{1}{*}{0.998} & \multirow{1}{*}{0} & \multirow{1}{*}{0.541} & \multirow{1}{*}{0.015} & \multirow{1}{*}{0.979}
\\[-0.225ex]
\clineB{1-10}{2}
%____________________________________________________________________________________
\end{tabular}
\end{center}
\end{table*}

\section{Empirical Validation} \label{sec:simulation}

We now compare the performance of the proposed method with recent state-of-the-art techniques: ISSM \cite{issm_matiwax_2022}, ADLLS \cite{zhang_adlls_2024}, and NBIC
\cite{ke_tsp_2021}. We consider a half-wavelength-spaced ULA with $M=64$ elements impinged upon by $r$ signals at $\phi_i^{}=-70^\circ_{}+\frac{i-1}{r-1}150^\circ_{}$, $i=1,\,\ldots,\,r$. The SNR of the $i$-{th} source is defined as $10 \log_{10}^{}\big( \sigma_{s_i}^2/\sigma_z^2 \big)$, where $ \sigma_z^2 = \frac{1}{M}\mathbb E\big[\norm{\mbf{z}(n)}_2^2\big]$. We consider two source settings: \textit{balanced sources}, where all sources have equal SNRs, and \textit{disparate sources}, where the source SNRs exhibit a linear trend over a $6\pspp$dB  span. We consider\footnote{A brief description of these models, along with results under Gaussian noise with diagonal covariance, is in the appendix.} (a) i.i.d Gaussian noise, (b) Gaussian noise with Banded Covariance structure (BCN) \cite{ssm_2021, issm_matiwax_2022} (c) heavy tailed heavy-tailed $\ell_2^{}$-noise with isotropic density function $f(\mbf{z}) \propto e^{- \zeta \lVert \mbf{z} \rVert_2^{}}\pspp$, $\zeta>0$, sampled according to \cite[Alg.  6]{kmech_jordan_2021}, and (d) \textit{Middleton class-A} Gaussian Mixture Noise (GMN) \cite{vastola1984threshold,chen2022nonparametric}.

The performance is studied under three different cases.
\begin{itemize}
    \item \textbf{Case 1:} $N=40$ snapshots are obtained from the array impinged by signals from $r=8$ sources at various SNRs. 
    \item \textbf{Case 2:} The number of sources $r$ is varied, and $N=40$ snapshots are obtained at $0\psp$dB SNR.
    \item \textbf{Case 3:} With $r=8$ sources, the number of snapshots is varied between $N=40$ and $N=80$ at $-3\psp$dB SNR.
\end{itemize}
We gauge the performance using the probability of correct detection (PCD) calculated as the fraction of 1000 random trials in which the number of sources is estimated correctly. 

Fig. \ref{fig:case1_results} depicts the results of Case 1.  We observe that the proposed method outperforms ADLLS, ISSM, and NBIC in different noise scenarios. Under i.i.d. Gaussian, $\ell_2^{}$-norm, and GMN, SEDAN achieves a PCD of 0.5 with about $5\pspp$dB and $6\pspp$dB lower SNR than ISSM and NBIC, respectively, and with $4\pspp$dB and $3\pspp$dB lower SNR under BCN. ADLLS fails here as the number of sources is too large for it to handle. The results for Case 2 are presented in Fig. \ref{fig:case2_results}. SEDAN performs significantly better than ADLLS, ISSM, and NBIC, particularly when the number of sources is very large. Specifically, our method  always achieves a high PCD of about 0.95 when $r$ is as large as 18, while ADLLS, NBIC, and ISSM totally fail when  the number of sources is above 10, 15, and 18, respectively. Table \ref{tab:case3_results} presents the performance of the proposed method in Case 3, with fewer $(N = 40)$ and more $(N = 80)$ snapshots than antennas $(M = 64)$. SEDAN clearly outperforms ADLLS, ISSM, and NBIC when $N<M$ and achieves a better or comparable performance to ADLLS when $N > M$.

\section{Conclusions} \label{sec:conclusion}

We presented a parameter-free source enumeration scheme, which (a) is able to handle both independent and colored Gaussian noise, as well as heavy-tailed noise,  (b) can handle a significantly larger number of sources than existing methods, and (c) can work with fewer snapshots. It harnesses the distribution of angles between high-dimensional points. We compared its performance with state-of-the-art methods through extensive simulations; in the challenging scenario where a high number of sources are to be estimated from a limited number of samples, our method significantly outperforms the existing ones. A comprehensive theoretical analysis of the proposed scheme could be considered as part of future work.

\appendices

\section{Additional Details on Noise Models}

The Gaussian noise with a banded covariance structure (BCN) models locally correlated  noise across antenna elements that are closely placed, and it has been considered in \cite{ssm_2021, issm_matiwax_2022}. The parameter $\gamma\in[0,1)$ controls the strength of local correlation: $\gamma = 0$ yields white Gaussian noise, while higher $\gamma$ increases neighboring correlation. We use the parameter $\gamma=0.5$ in our analysis.

\textit{Middleton Class-A} noise models heavy-tailed noise with a mixture of Gaussians (GMN) \cite{vastola1984threshold}. It is commonly considered in the context of signal detection with impulsive interference \cite{chen2022nonparametric}. The parameter $A>0$ controls the impulsiveness, while $\Gamma>0$ represents the ratio of the background Gaussian to impulsive power. As $A \to \infty$ or $\Gamma \to \infty$, GMN corresponds to Gaussian noise. We use the truncated variant with 10 Gaussian components and the parameters $A=0.3$ and $\Gamma = 0.005$ following \cite{chen2022nonparametric}.

In addition, we also consider the Gaussian noise with a diagonal covariance structure (DCN) \cite{ssm_2021,issm_matiwax_2022} in the following section. It models uncorrelated noise with unequal power across antennas and is often used in scenarios involving antenna-specific variations. The parameter $\beta\in[0,1)$ controls the degree of power variation: $\beta = 0$ yields white Gaussian noise, while higher $\beta$ increases disparity. In our analysis, we consider $\beta=0.5$.

\begin{figure*}[!b]
\centering
    \begin{minipage}[b]{0.495\textwidth}
        \centering
        \includegraphics[width=.865\linewidth]{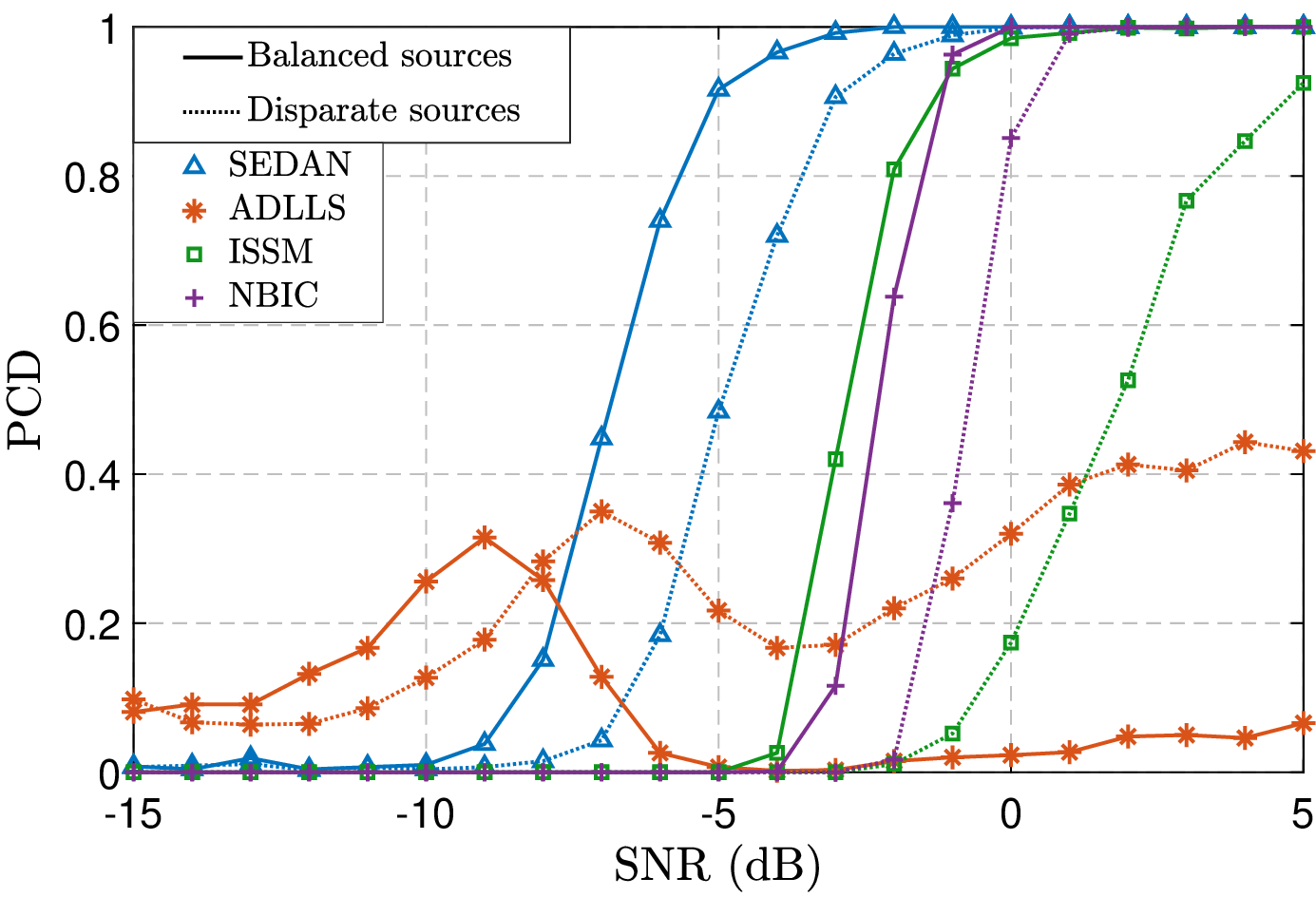}
        \caption{PCD for Case 1 under DCN.}
        \label{fig:DCN_results1}
    \end{minipage}
    \hfill
    \begin{minipage}[b]{0.495\textwidth}
        \centering
        \includegraphics[width=.865\linewidth]{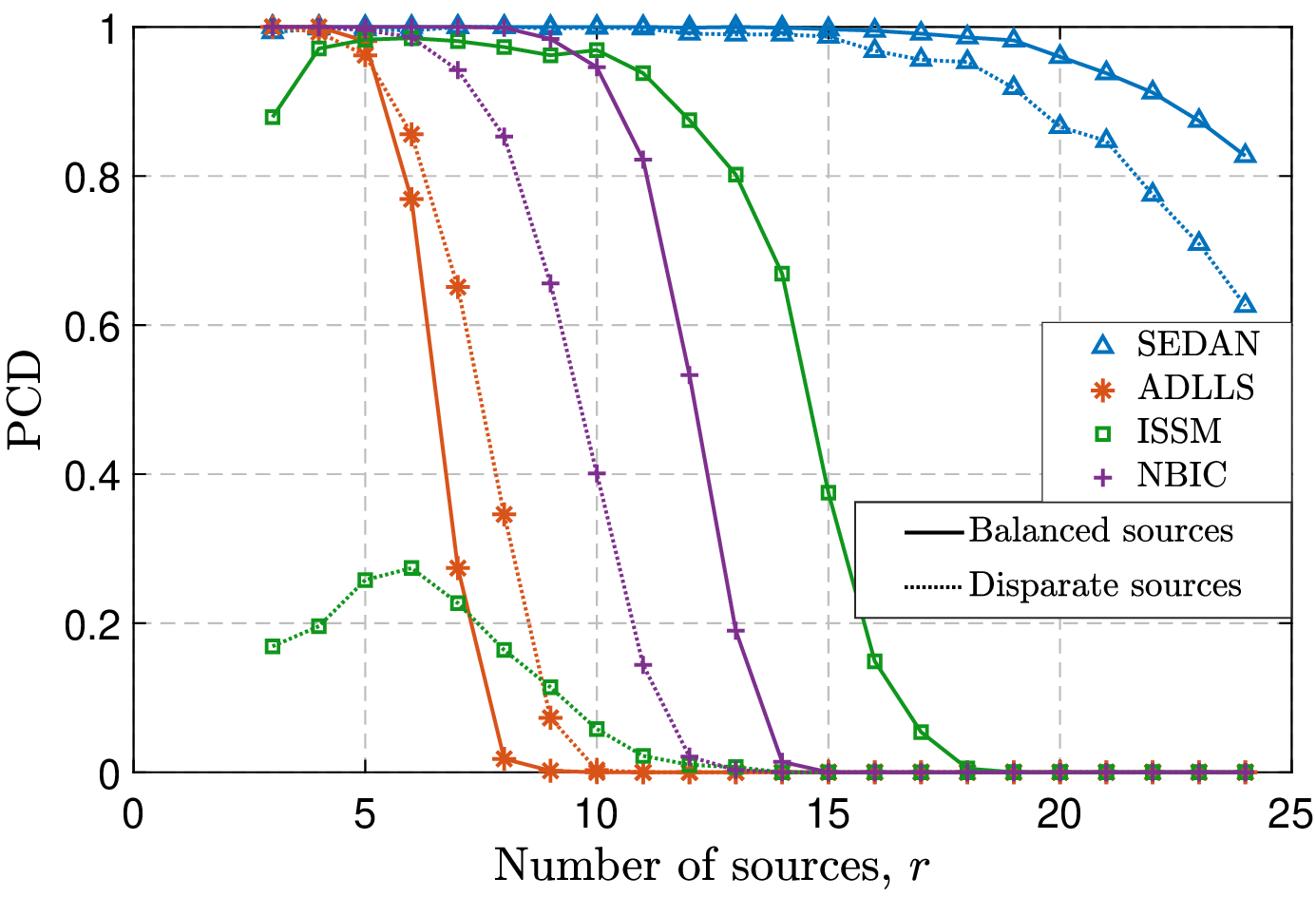}
        \caption{PCD for Case 2 under DCN.}
        \label{fig:DCN_results2}
    \end{minipage}
\end{figure*}

\section{Performance under Diagonal Covariance Noise} \label{appn:dcn_results}

In this section, we evaluate the performance of the proposed method under Gaussian noise with a diagonal covariance structure (DCN), which models unequal noise powers across the antenna elements. We adopt the parameter setting $\beta = 0.5$, following the configurations in \cite{issm_matiwax_2022} and \cite{zhang_adlls_2024}. As detailed in Section \ref{sec:simulation}, the method is tested under three different cases with $M=64$ antennas. The results of Case 1 and Case 2 are depicted in Figs. \ref{fig:DCN_results1} and \ref{fig:DCN_results2}, respectively, and the results of Case 3 are provided in Table \ref{tab:DCN_results}.

\begin{table}[h!]
\setlength{\tabcolsep}{5pt}
\renewcommand{\arraystretch}{0.91}
\caption{PCD for Case 3 under DCN.}
\label{tab:DCN_results}
\begin{center}
\begin{tabular}{? C{4.25em} *{2}{? C{4em} | C{4em}} ?} 

\clineB{2-5}{2}

\multicolumn{1}{c?}{}  &  \multicolumn{2}{c?}{\multirow{1}{*}{\textbf{Balanced sources}}} &  \multicolumn{2}{c?}{\multirow{1}{*}{\textbf{Disparate sources}}} 
\\[-0.225ex]
\cline{2-5}

\multicolumn{1}{c?}{} & \multirow{1}{*}{${N}\nsp=\nsp40$} & \multirow{1}{*}{${N}\nsp=\nsp80$} & \multirow{1}{*}{${N}\nsp=\nsp40$} & \multirow{1}{*}{${N}\nsp=\nsp80$} 
\\[-0.225ex]
\clineB{2-5}{2}

\specialrule{0.8pt}{2pt}{0pt}
%____________________________________________________________________________________
\multirow{1}{*}{SEDAN} & \multirow{1}{*}{\textbf{0.993}} & \multirow{1}{*}{\textbf{1}} & \multirow{1}{*}{\textbf{0.889}} & \multirow{0.997}{*}{\textbf{0.997}}
\\[-0.225ex]
\cline{1-5}

\multirow{1}{*}{ADLLS} & \multirow{1}{*}{0.004} & \multirow{1}{*}{\textbf{1}} & \multirow{1}{*}{0.163} & \multirow{1}{*}{\textbf{1}}
\\[-0.225ex]
\cline{1-5}

\multirow{1}{*}{ISSM} & \multirow{1}{*}{0.408} & \multirow{1}{*}{0.981} & \multirow{1}{*}{0} & \multirow{1}{*}{0.034}
\\[-0.225ex]
\cline{1-5}

\multirow{1}{*}{NBIC} & \multirow{1}{*}{0.115} & \multirow{1}{*}{\textbf{1}} & \multirow{1}{*}{0} & \multirow{1}{*}{0.891}
\\[-0.225ex]
\clineB{1-5}{2}
%____________________________________________________________________________________
\end{tabular}
\end{center}
% \vspace{-2ex}
\end{table}

\begin{figure*}[h!]
\begin{minipage}[b]{.5\textwidth}
\centering
\includegraphics[width=.925\linewidth]{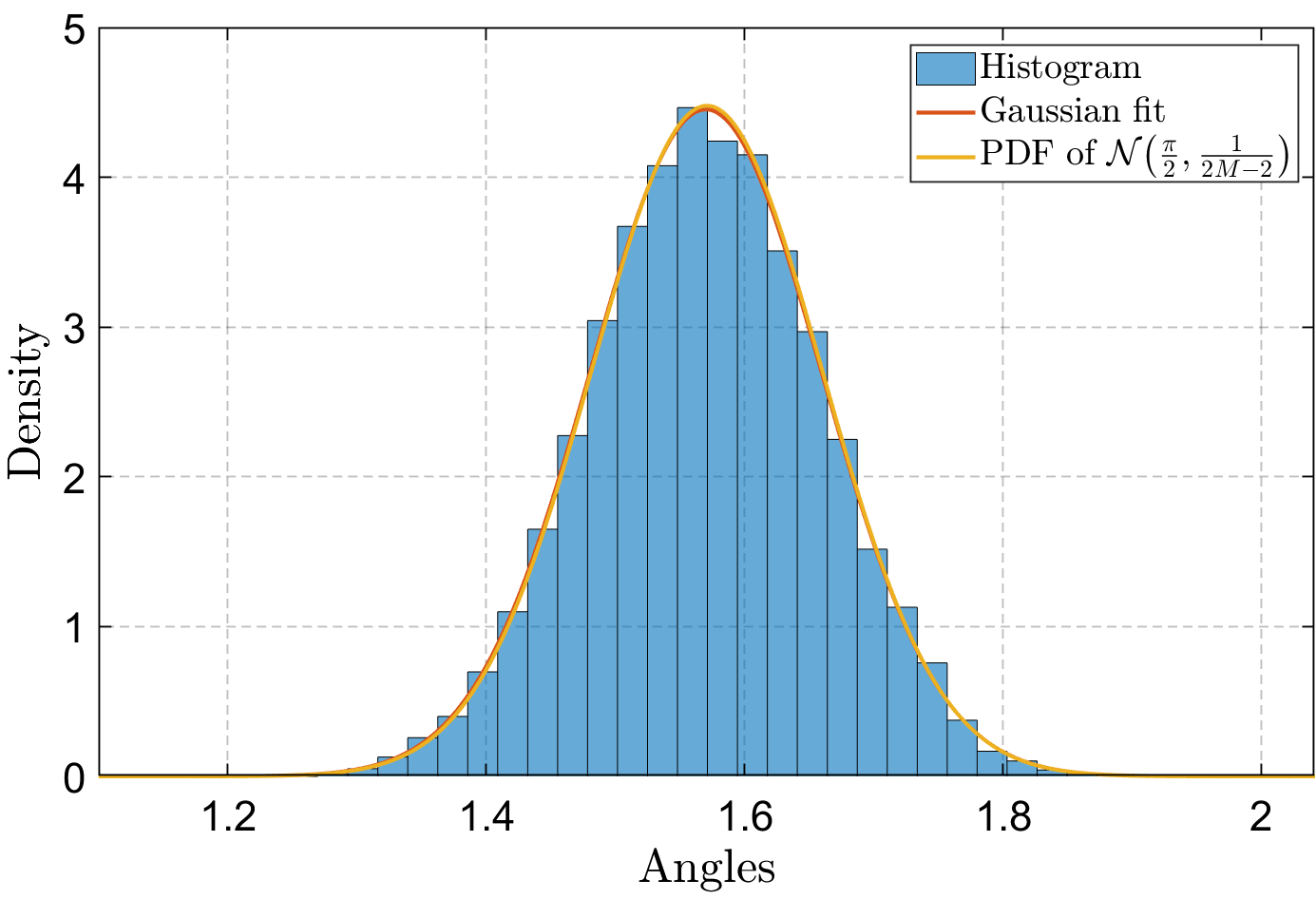}
\caption{Distribution of angles under i.i.d. Gaussian noise.%
\\[5ex]
}
\label{fig:Hist_M64_N1000_White}
\end{minipage}
\begin{minipage}[b]{.5\textwidth}
\centering
\includegraphics[width=.925\linewidth]{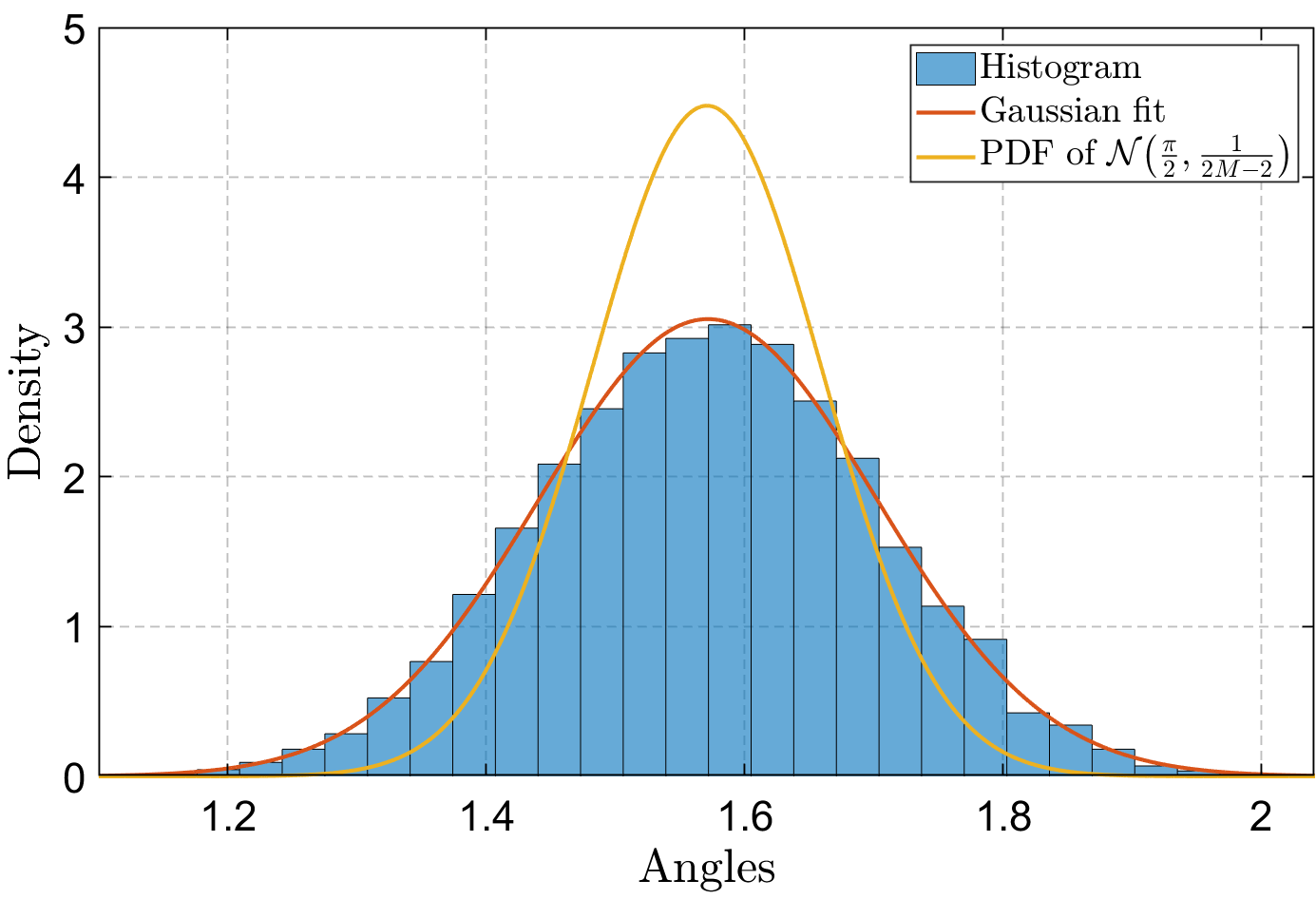}
\caption{Distribution of angles under BCN.%
\\[5ex]
}
\label{fig:Hist_M64_N1000_Banded}
\end{minipage}
\begin{minipage}[b]{.5\textwidth}
\centering
\includegraphics[width=.925\linewidth]{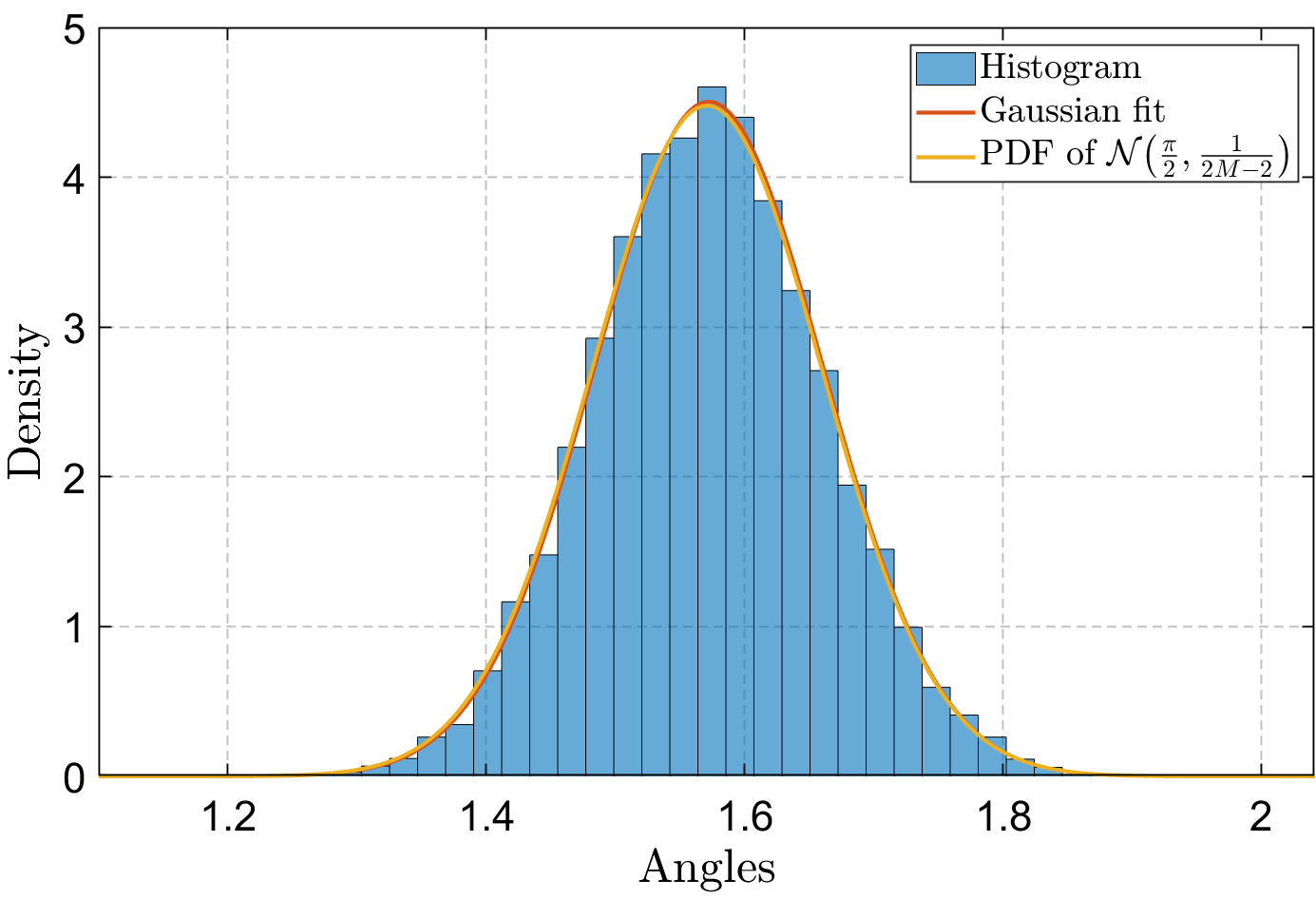}
\caption{Distribution of angles under $\ell_2^{}$-noise.%
\\[5ex]
}
\label{fig:Hist_M64_N1000_2norm}
\end{minipage}
\begin{minipage}[b]{.5\textwidth}
\centering
\includegraphics[width=.925\linewidth]{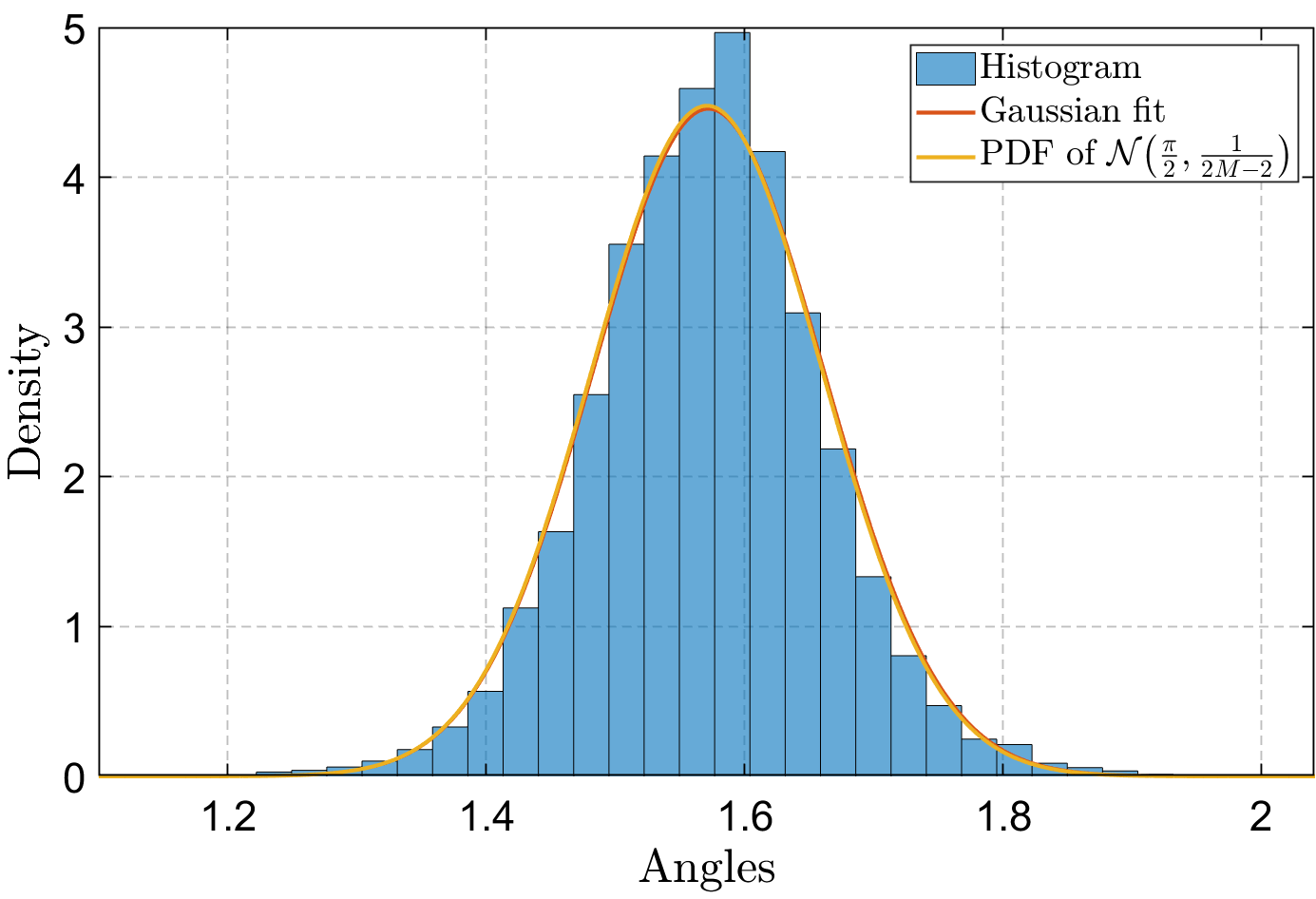}
\caption{Distribution of angles under GMN.%
\\[5ex]
}
\label{fig:Hist_M64_N1000_GMN}
\end{minipage}
\begin{center}
\begin{minipage}[b]{.5\textwidth}
\centering
\includegraphics[width=.925\linewidth]{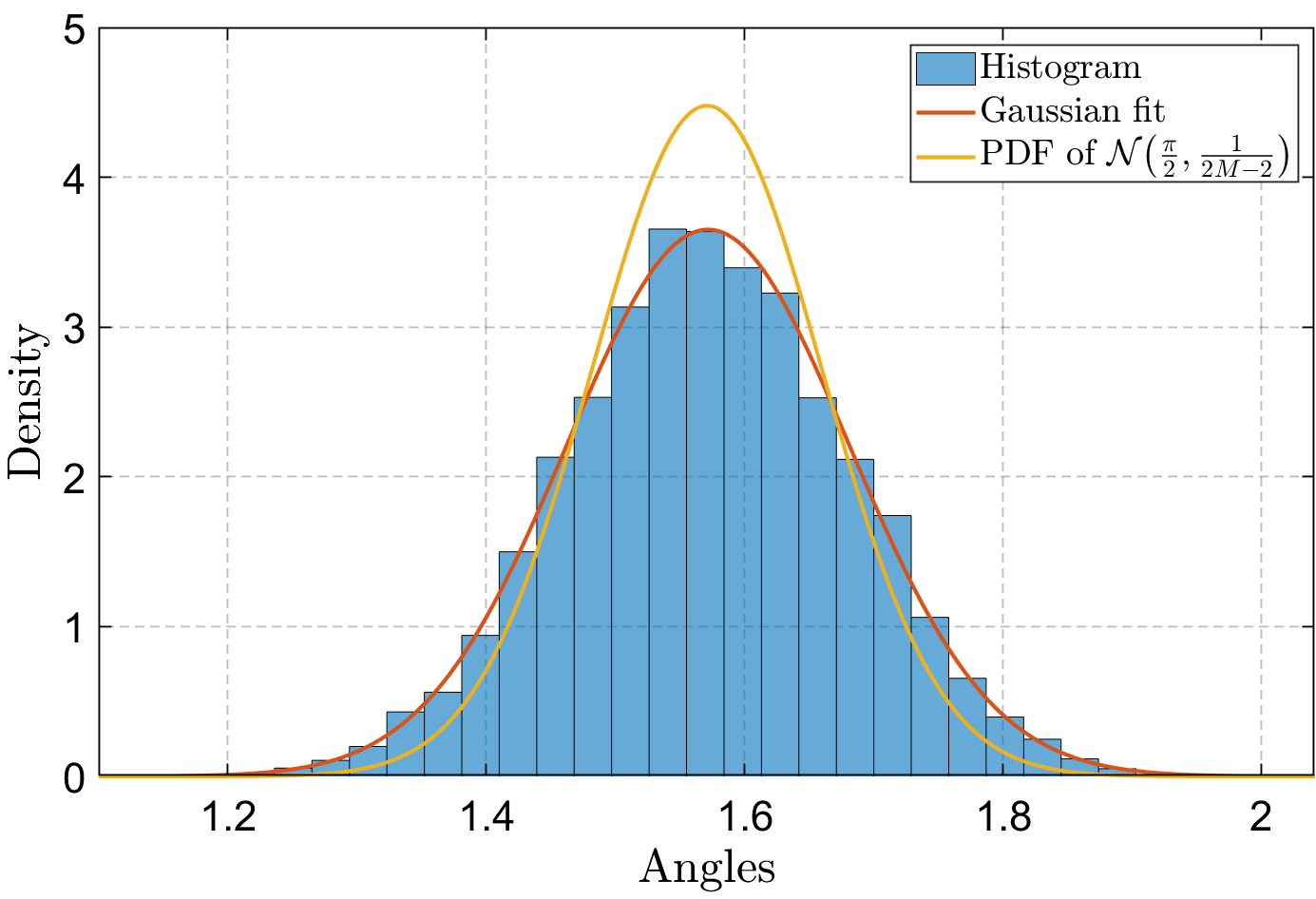}
\caption{Distribution of angles under DCN.%
% \\[-6.5ex]
}
\label{fig:Hist_M64_N1000_DCN}
\end{minipage}
\end{center}
\vspace{-0.65ex}
\end{figure*}

Even in this scenario, the proposed method demonstrates superior performance compared to the state-of-the-art, despite the fact that the noise density of DCN is non-isotropic. Specifically, as shown in Fig. \ref{fig:DCN_results1}, SEDAN achieves a PCD of 0.5 with about $4\pspp$dB and $5\pspp$dB lower SNR compared to ISSM and NBIC, respectively, while ADLLS fails drastically in this setting. Also from Fig. \ref{fig:DCN_results2}, it can be seen that the proposed method sustains a high PCD of 0.95 even with 20 sources, while ADLLS, NBIC, and ISSM completely fail when the number of sources is above 10, 15, and 18, respectively. Additionally, Table \ref{tab:DCN_results} highlights the effectiveness of SEDAN in challenging scenarios involving \textit{disparate sources} with fewer snapshots than the number of antennas, where it achieves a PCD of 0.88, in contrast to the other methods, which fail to achieve even a PCD of 0.2.

\section{Distribution of Angles under Various Noise Models} \label{appn:emp_dist_angles}

The proposed method leverages the statistical distribution of angles between the high-dimensional points. Specifically, Lemma \ref{lemma:gaussian} establishes that the angle subtended by the noise vectors from an isotropic distribution approximately follows the Gaussian distribution with mean $\frac{\pi}{2}$ and variance $\frac{1}{2(M-1)}$. However, to extend the applicability of our method to non-isotropic noise settings, we assume that the angles subtended by the noise vectors follow a Gaussian distribution with an arbitrary mean and variance. We now empirically validate this assumption, analyzing the distribution of angles under various noise models considered in our study. 

Considering an array with $M = 64$ antennas, the empirical distributions of the angles between noise vectors are computed and compared with the corresponding Gaussian fits, as well as the density function of $\mathcal{N}\big(\frac{\pi}{2} , \frac{1}{2(M-1)}\big)$, and the results are plotted in Figs. \ref{fig:Hist_M64_N1000_White}-\ref{fig:Hist_M64_N1000_DCN}. It is evident that the angles are normally distributed under all the noise models considered. In the case of isotropic noise\textemdash specifically, i.i.d. Gaussian and $\ell_2^{}$-norm noise\textemdash the angles precisely follow $\mathcal{N}\big(\frac{\pi}{2} , \frac{1}{2(M-1)}\big)$, validating Lemma \ref{lemma:gaussian}. For other noise models, namely, BCN, DCN, and GMN, despite the noise densities being non-isotropic, the empirical distribution of angles between noise vectors closely matches a Gaussian distribution with an arbitrary mean and variance.

This demonstrates that the proposed method remains valid even under noise conditions that extend beyond the assumptions of Lemma 1. These results justify the applicability of the proposed method under broader noise conditions than those assumed in Lemma \ref{lemma:gaussian} and explain its robust performance under heavy-tailed, as well as colored noises.

\bibliographystyle{IEEEtran}
\bibliography{refs}

\end{document}